\pdfoutput=1
\documentclass[a4paper,12pt]{article}
\usepackage[utf8]{inputenc}
\usepackage{amssymb,amsmath,amsfonts,bbm,dsfont,graphicx,hyphenat,bm,float,a4wide,xcolor}
\usepackage[english]{babel}
\usepackage[colorlinks,linkcolor=blue,citecolor=blue,urlcolor=blue]{hyperref}
\usepackage[square,numbers,sort&compress]{natbib}
\usepackage[caption=false]{subfig}
\usepackage[normalem]{ulem}
\usepackage{etoolbox}
\AtBeginEnvironment{thebibliography}{\small}

\newcommand{\beq}{\begin{eqnarray}}
	\newcommand{\eeq}{\end{eqnarray}}

\DeclareMathOperator{\tr}{tr}

\newcommand{\bea}{\begin{eqnarray}}
	\newcommand{\eea}{\end{eqnarray}}
\newcommand{\be}{\begin{equation}}
	\newcommand{\ee}{\end{equation}}

\def\nn{\nonumber}

\begin{document}
	\title{
 \vskip 20pt
 \bf{Confinement by  Monopole Loops  in  Inhomogeneous Magnetic Field}
	}
	\vskip 40pt 
	
	\author{
S.~Bolognesi
 \\[2pt]
 {\em \footnotesize
 Department of Physics ``E. Fermi", University of Pisa, and INFN, Sezione di Pisa}\\[-5pt]
{\em \footnotesize
	Largo Pontecorvo, 3, Ed. C, 56127 Pisa, Italy}\\[2pt]
{\footnotesize \texttt{stefano.bolognesi@unipi.it}}	}
	
	\date{}
	
	\vskip 10pt
	\maketitle

 \begin{abstract}
We show that a generalized Polyakov mechanism can lead to confinement at weak coupling in $3+1$ dimensions when the theory is placed in a non‑trivial, spatially varying magnetic field background. Depending on the magnitude of the field and the length scale of its spatial variation, the “dual” Schwinger mechanism for monopole-antimonopole pair creation may or may not be operative. At the threshold, monopole loops in the Euclidean description develop an almost flat direction. In this regime, confinement arises in a way similar to the $2+1$ dimensional Polyakov mechanism and the monopoles and antimonopoles are effectively replaced by deconfined “bits” of a monopole loop.
\end{abstract}
 
 \newpage

 We consider a mechanism of confinement assisted by a background magnetic field. We take a weakly coupled $3+1$-dimensional theory that has a massless $U(1)$ gauge field and admits massive magnetic monopoles. This is the standard Georgi–Glashow model which, when considered in $2+1$ dimensions, realizes confinement via the Polyakov mechanism \cite{Polyakov:1976fu}. In the Euclidean formulation the monopole–antimonopole gas screens the magnetic field and gives a mass to the dual photon. 
 We consider the same theory in $3+1$, or $4$ Euclidean, dimensions. Now monopoles are strings, there is no localized finite-action instanton solution, and thus the Polyakov mechanism does not work. Here we show that, in a particular inhomogeneous magnetic-field background, one can apply ideas similar to Polyakov’s confinement and obtain confinement in $3+1$ that also holds at weak coupling. 
 
 Many works have explored ways to connect to or use the Polyakov mechanism for $4$D theories;  an example of adiabatic continuity is \cite{Shifman:2008ja}, but there are many other realizations. Here we present a mechanism that can work in $4$D without the need for compactification, at weak coupling. The essential ingredient is the spatial variation of the background magnetic field, which provides a mechanism to deconfine monopole loops and thus generate a mass gap for the dual photon. The deconfinement of monopole strings has been suggested before \cite{Polyakov:2004vp} and used in $5$D in \cite{Bolognesi:2011rq}. Other works on monopole loops and confinement include \cite{Nguyen:2025voy}. Often the deconfinement of monopole strings is associated with a Hagedorn phase transition; here we use a different mechanism. A relation between Polyakov confinement mechanism in $3$D and dual superconductivity with monopole condensation in $4$D has been studied in \cite{Auzzi:2008zd,Dvali:2007nm} for the case of a domain wall as a dual Josephson junction. 
 The work we will present can be considered as yet another relation between the two mechanisms of confinement.

 Pair production in inhomogeneous backgrounds has been intensively studied in the context of the Schwinger effect in electrodynamics, for example  \cite{Brezin:1970xf,Schutzhold:2008pz}. One of the main motivations is that time variation, especially, can enhance pair production and thus make the effect stronger and hopefully measurable (e.g. in strong electromagnetic-wave backgrounds). Spatial variations instead tend to lower the pair-production probability. This is the effect we consider in the present paper. For spatially varying background fields, unlike in the constant case, there is in general a critical value of the field below which the vacuum is stable and no pair production occurs. Here we are interested precisely in what happens at the critical value for magnetic monopoles. The dual version of  Schwinger effect, that is the pair production of monopole-antimonopole pairs by a background magnetic field, was first discussed in \cite{Affleck:1981ag,Affleck:1981bma} and in holographic context in \cite{Bolognesi:2012gr}.

 The theory we consider is an $SU(2)$ gauge theory with an adjoint field $\Phi$ in $4$ dimensions. We take the Euclidean action in $x,y,z,\tau$
 \beq
 S_{4E} = \int d^4x \left(\frac{1}{2g^2} \tr F^2 + \frac{1}{2} \tr (D\Phi)^2 + \lambda \Big( \tr \Phi^2 - \frac{v^2}{2} \Big)^2 \right) \ .
 \label{action4D}
 \eeq
 The vacuum $\Phi = \frac{1}{2}{\rm diag} (v,-v)$ breaks $SU(2) \to U(1)$.
 The mass of the $W$-boson is $m_W = g v$. The mass of the ’t Hooft–Polyakov monopole \cite{tHooft:1974kcl,Polyakov:1974ek} is 
 \beq
 m_{M} = \alpha \frac{1}{g v} = \alpha \frac{m_W}{g^2}
  \ .
  \eeq
  For reference we can take the BPS case $\lambda =0$ for which $\alpha = 4\pi$. The abelian gauge field is 
 \beq
 a_{\mu} = \frac{2}{v} \tr (A_{\mu} \Phi) \ .
 \eeq
 The abelian magnetic flux of the monopole or antimonopole is $\Phi_B = \pm 4\pi$. If we compactify one direction with periodic boundary conditions, with the identification $z \equiv z + L$, we obtain the $2+1$ model of the low-energy modes
 \beq
 S_{3E} = \int d^3x \left(\frac{1}{2e^2} \tr F^2 + \frac{1}{2} \tr (D\phi)^2 + \lambda \Big( \tr \phi^2 - \frac{w^2}{2} \Big)^2 \right)\ ,
 \label{effective3D}
 \eeq
 where the correspondence of the parameters is
 \beq
 L e^2 = g^2 \ , \qquad \phi = \sqrt{L} \Phi \ , \qquad  w = \sqrt{L} v \ .
 \eeq
 This effective action (\ref{effective3D}) neglects the Kaluza–Klein modes with mass $\propto \frac{1}{L}$.
 The mass of the $W$-boson remains the same, $m_W = e w = g v$. The monopole-instanton corresponds to the $4$D monopole string wrapped around the compactified direction and has action $S_{M} = m_{M} L = \alpha \frac{m_W}{e^2}$. Henceforth we refer to $m_{M}$ as the mass of the monopole in $4$D and to $S_{M}$ as the action (a dimensionless quantity) of the monopole-instanton in $3$D.

 Let us take $L$ finite and apply the Polyakov mechanism of confinement to the effective action (\ref{effective3D}); this works if all the relevant length scales (i.e. the monopole-antimonopole average distance and the inverse of the confinement mass gap) are larger than $L$.  When a length scale becomes of order $L$ or smaller we enter a $4$D regime and some aspects of the computation must be modified (in principle this is actually good: it means the confinement mechanism survives to $4$D). The liberation of monopoles implies that the system is described by a statistical ensemble of monopoles and antimonopoles Coulomb gas.
 The instanton (in this case a monopole) measure accounts for the exponential suppression due to the instanton action and a prefactor due to integral measure for the zero modes (massive modes also give a  contribution but we neglect them). The prefactor is quite important for us, so we consider it explicitly. We neglect instead contributions from massive modes. The zero modes of a single monopole are parametrized by $\mathbb{R}^3 \times S^1$. The metric for the zero modes of a physical monopole is $g_{AB} = S_{inst} \, \delta_{AB}$, where $A,B = \Big(x,y,\tau,\frac{\theta}{m_W}\Big)$ where the  $S^1$, parameterized by $\theta \in (0,2 \pi)$, has physical radius $\frac{1}{m_W}$. The instanton measure is proportional to the $\sqrt{{\rm det} \, g_{A B}} e^{-S_{inst}}$ but also it needs a UV regulator. We are following  here the derivation  of \cite{Shifman:2012zz}. Each translation zero mode contributes a prefactor $\propto \sqrt{S_{inst}} m_{UV}$, where $m_{UV}$ is a regularization scale which we always take $\sim m_W$. There is also the angular variable for the global $U(1)$ gauge transformation with a prefactor $\propto \sqrt{S_{inst}} m_{UV}\frac{1}{m_{W}}$, which we integrate over. The measure in $\mathbb{R}^3$, neglecting all massive modes around the monopole-instanton,  is then
 \beq
 d\mu^{(3)} = \xi^{(3)} dx dy d\tau \ , \qquad \quad \xi^{(3)} \propto \Big(\sqrt{S_{inst}} m_{UV}\Big)^4 \frac{1}{m_{W}} e^{-S_{inst}}  \ ,
 \eeq
 with
 \beq
 S_{inst} = S_{M} =\frac{\alpha m_{W}}{e^2}\ , \qquad m_{UV} \simeq m_W \ .
 \eeq
 The density of the  monopole–antimonopole gas is
 \beq
 \xi^{(3)} = \frac{1}{\lambda_{gas}^3 } \simeq \frac{m_{W}^5}{e^4} e^{-\frac{\alpha m_{W}}{e^2}} \ ,
 \eeq
 where $\lambda_{gas}$ sets the average separation. The dual photon in $3$ dimensions is a scalar field
 \beq
 \frac{1}{e} f_{\mu\nu} = e \epsilon_{\mu\nu\rho} \partial_{\rho} \varphi \ .
 \eeq
 The low-energy $U(1)$ action with the interaction with monopoles is most conveniently written with the dual photon
 \beq
 S_E = \int d^3x \left( \frac{e^2}{2} (\partial \varphi)^2 + V(\varphi) \right) \ .
 \label{actionPhi}
 \eeq
 Considering only one monopole or antimonopole contribution (dilute gas approximation) we have
 \beq
 V(\varphi) \simeq \xi^{(3)} (e^{i \varphi} + e^{-i\varphi}) =
 2 \xi^{(3)} \cos{\varphi} \ .
 \label{potPhi}
 \eeq 
 For small field,
 \beq
 V(\varphi) \simeq \frac{1}{2}e^2  M_{\varphi}^2  \varphi^2
 \eeq
 from which the mass of the dual photon 
 which sets the confinement  scale 
 \beq
 M_{\varphi}^2 = \frac{1}{\lambda_{conf}^2} =\frac{2 \xi^{(3)}}{e^2} \ .
 \eeq
 The mass of the dual photon is also the inverse of the thickness of the confining string $\lambda_{conf}$. The confining string is the sine-Gordon domain wall of the action (\ref{actionPhi}),(\ref{potPhi}). At the end, the two important length scales in the problem are 
 \beq
 \lambda_{gas} \simeq \frac{e^{4/3}}{m_{W}^{5/3}} e^{\frac{\alpha m_{W}}{3 e^2}} \ ,\qquad \lambda_{conf} \simeq \frac{e^{3}}{m_{W}^{5/2}} e^{\frac{\alpha m_{W}}{2 e^2}} \ .
 \label{scaleswithoutwalls}
 \eeq
 The approximation works if the following hierarchy between the length scales is satisfied
 \beq
 \lambda_{conf} \gg \lambda_{gas} \gg L \gg \frac{1}{m_W} \ .
 \eeq
 These conditions are achievable at weak coupling. The approximation requires a dilute gas of monopole-instantons and a confinement scale even larger than the average monopole separation. Note that confinement may also occur at strong coupling or for a non-dilute gas of monopoles and antimonopoles, but in that case the computation is much more complicated.
 
 Now if we increase $L$ without altering the background, the formula shows that the confinement length scale grows
 exponentially with $L$, due to the increase in the monopole mass. This is why the Polyakov mechanism of confinement, as it stands, “disappears” in the decompactification limit as  $\frac{\lambda_{conf}}{L} \to \infty$ exponentially fast as $L$ grows. To evade this we will put the theory in a particular background and redo the computation. First we need to discuss the pair production of monopoles in a magnetic background.

 It is known that a constant magnetic field triggers monopole pair production. A physical monopole–antimonopole pair can be created out of the vacuum if the background field is strong enough. This costs an energy $2 m_{M}$, but at the distance $\frac{2 m_{M}}{B}$ the monopoles are already beyond the tunneling barrier. This is the dual version of the Schwinger effect of electrically charged particle–antiparticle (for example electron-positron in QED) pair production in an external electric field. At leading order the tunneling probability is $e^{- \frac{\pi m_{M}^2}{B}}$. From the Euclidean perspective this tunneling process is described by a ``bounce'' solution which, for a constant magnetic-field background, is a circular monopole loop of radius $\frac{m_{M}}{B}$. It is similar to the vacuum decay in a metastable potential. 
 
 The worldline description of monopole-instanton pair production is dominated, in the semiclassical regime, by the solution of the Euclidean worldline action
 \beq
 S_{E} = \int dl \,  m_{M} \sqrt{\dot{x}^{\mu} \dot{x}^{\mu}} \pm \int d l \, \dot{x}^{\mu} \tilde{a}^{\mu} \ ,
 \label{worldlineaction}
 \eeq
 where $l$ is a parametrization of the worldline, $\dot{x}^{\mu} = \frac{dx^{\mu}}{dl}$ and $\tilde{a}_{\mu}$ is the dual abelian gauge field. The sign $\pm$ reflects the choice of orientation of monopole or antimonopole, or equivalently the choice of orientation of the monopole loop.  A closed monopole loop thus has an action that depends on just two geometric quantities
 \beq
 S_E = m_{M} P \pm \Phi_B \ ,
 \eeq
 where $P$ is the perimeter of the loop and $\Phi_B$ is the magnetic flux through it. Taking $\vec{B}$ along a constant direction, say $z$, we can restrict to loops in the $(\tau,z)$ plane in order to minimize the perimeter for a given flux.  For the pair-production problem we need a stationary solution, so the overall sign of $\pm \Phi_B$ must be negative. We use an affine parametrization where $l \in [0,1]$ with the ends identified, the norm of the velocity is conserved
 \beq
 \label{constderqua}
 \dot{x}^{\mu}\dot{x}^{\mu} = {\rm const} = P^2 \ ,
 \eeq
 The equation is then 
 \beq
 \frac{m_{M}\ddot{x}^{\mu} }{ P } = \pm \tilde{f}_{\mu\nu}\dot{x}^{\nu} \ .
 \eeq
 For a constant field $\tilde{f}_{z \tau}=B$ the solution is a circular trajectory
 \beq
 z(l) =\frac{m_{M}}{ B} \cos{(2\pi l)} \ ,  \qquad \tau(l) = \frac{m_{M}}{B} \sin{(2\pi l)} \ ,
 \eeq
 for which  the action is $ S_E = \frac{\pi m_{M}^2}{B}$. 
 This is indeed the local maximum of the action evaluated on generic circle of radius $R$
 \beq
 S_E = m_{M} 2 \pi R - B \pi R^2 
 \eeq
 which is attained at $R_{*} = \frac{m_M}{B}$. From the point of view of $R$ this is a maximum, but in general is a saddle point when considering all other deformations of the loop.

\begin{figure}[h]
 	\centering
 	\includegraphics[width=0.6\linewidth]{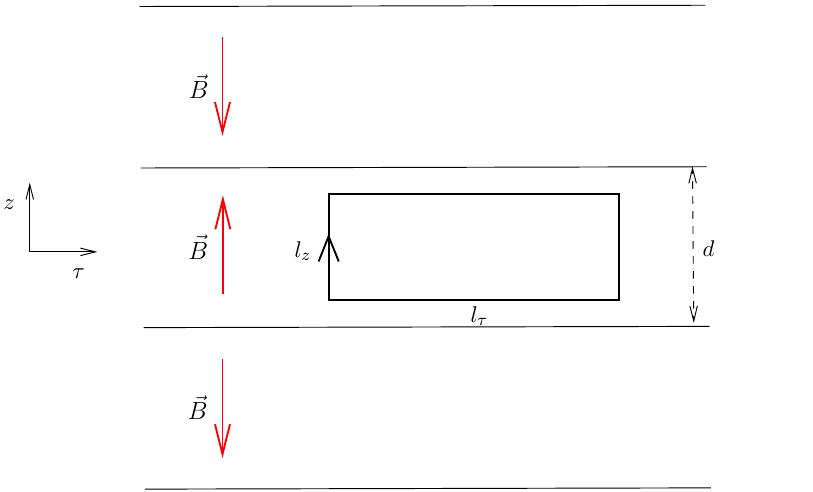}\\	
 	\caption{\small Alternate strips of constant magnetic field and a rectangular monopole loop centered in the middle on a strip  with side lengths $l_{z}$ and $l_{\tau}$. 	}
 	\label{ltlz}
 \end{figure}

 Now we come to an inhomogeneous background field.  The worldline formalism was used for the Schwinger effect in inhomogeneous backgrounds in \cite{Dunne:2005sx}, and for strings in \cite{Bolognesi:2016qme}. The same technique can be used for the dual Schwinger effect.  Here we are mostly interested in a spatially varying background magnetic field, and the simplest case to consider is the alternating magnetic-field background with a shape that may depend on the case. Let us for the moment approximate the magnetic field $\vec{B} = B \hat{z}$ as a constant inside a segment $z \in (0,d)$ and then alternating in sign (see Figure \ref{ltlz}) so that it is $2d$ periodic.  The monopole loop solution is in a plane at fixed $(x,y)$, when  $z$ is  the direction of the magnetic field and $\tau$ the Euclidean time.  
 For sufficiently large $B$ the bounce is smaller than $d$, $B > B_{cr} = \frac{2 m_{M}}{d}$, and thus pair production proceeds essentially as in a constant  background. For smaller $B< B_{cr}$ there is instead no pair production: the spatial variation does not allow a closed monopole loop to solve the equations. 
%

  To understand better the problem let us approximate the loop by a rectangle with side lengths $l_{z}$ and $l_{\tau}$, centered in the middle of the strip where the magnetic field is constant, see Figure \ref{ltlz}. The strip has width $d$ and inside the magnetic field is constant. Then the magnetic field repeats with alternating signs. We then find 
 \beq
 S_E = 2 m_{M} (l_z + l_{\tau}) - B l_{\tau} f(l_z) \ ,
 \eeq
with \beq  f(l_z) = 
 \left\{\begin{array}{lll} l_z  & {\rm if} & l_z < d \ , \\  2d - l_z &  {\rm if} &  d<l_z<3d \ , \\ 
 	  -4d  + l_z &  {\rm if} &  3d<l_z<4d \ , \\ 
 	 \end{array} \right. \qquad \quad f(l_z) = f(l_z + 4 d)  \ .
 \eeq 
 \begin{figure}[h]
 	\centering
 	\includegraphics[width=0.32\linewidth]{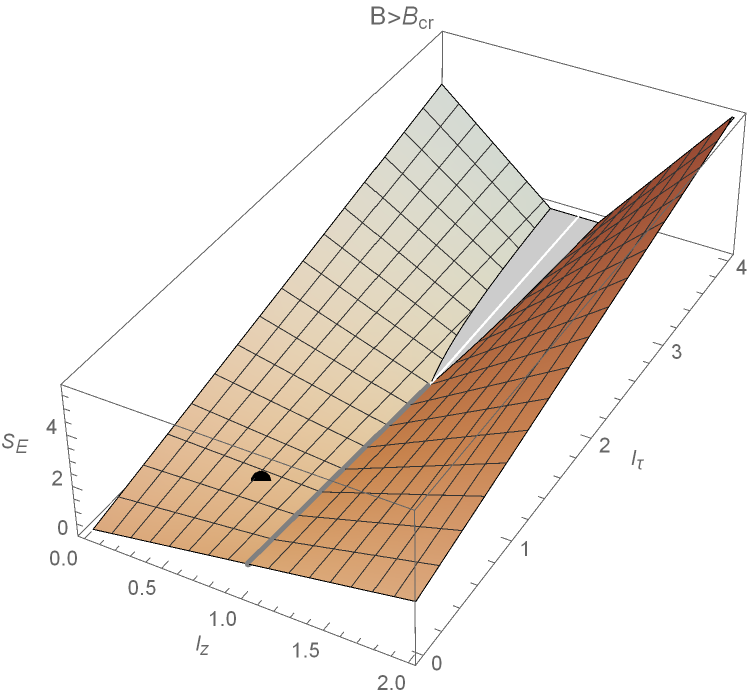}
 	\includegraphics[width=0.32\linewidth]{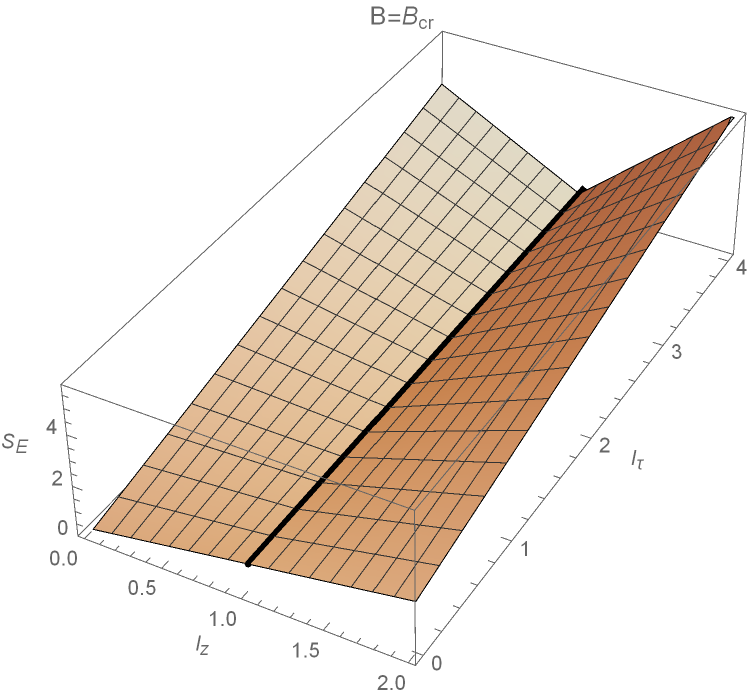}
 	\includegraphics[width=0.32\linewidth]{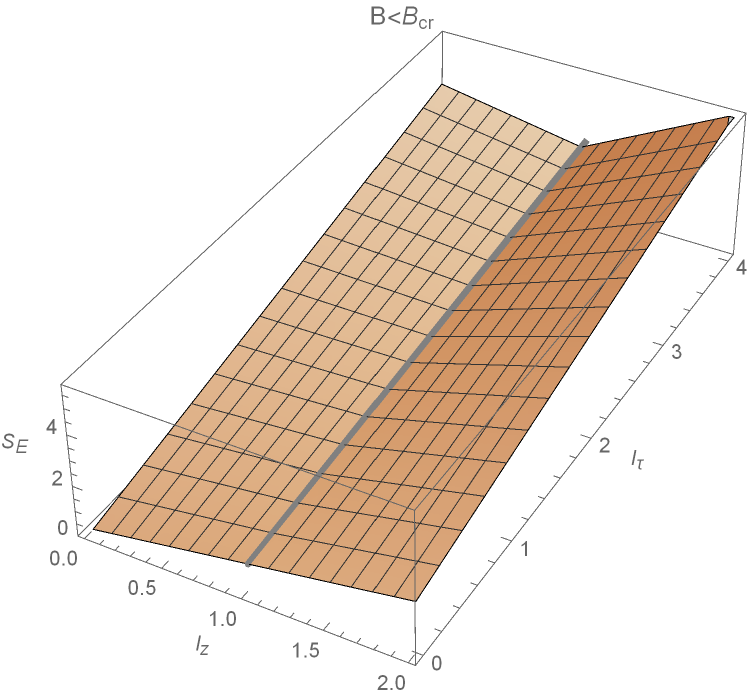}
 	\caption{\small The Euclidean action $S_E(l_{\tau},l_{z})$ for $B$ above, equal to, and below the critical value of $B_{cr}$. For $B>B_{cr}$ there is a saddle point solution (black dot in the left  plot).   At the critical value $B=B_{cr}$ there is a flat direction (black line in the middle plot). Here we used $m_{M}=.5$ and $d=1$.
 	}
 	\label{lpl}
 \end{figure}
 See Figure \ref{ltlz} for plots of $S_E(l_{\tau},l_{z})$.
 For $B>B_{cr}$ there are directions of instability: at fixed $d>l_z>l_{z,cr}$ and for large $l_{\tau}$ we have $S_E = 2 m_{M} l_z + ( 2 m_{M} - B l_z ) l_{\tau}$, which decreases linearly to $-\infty$ as $l_{\tau} \to \infty$. The case $B = B_{cr}$ is special because there is no instability: for $l_z = d$ we have a flat direction, $S_E = 2 m_{M} d$, for any $l_{\tau}$ (for a general case it will be an almost flat direction).
 This monopole loop is thus an instanton with a flat direction $l_{\tau}$. A section at fixed $z$ crosses the loop at two points corresponding to a monopole and an antimonopole in $\mathbb{R}^3$; they can then be created at any separation $l_{\tau}$. In this sense we can say that at $B= B_{cr}$ we have a sort of deconfinement of the monopole loop.

\begin{figure}[htp]
	\centering	
	\includegraphics[width=0.55\linewidth]{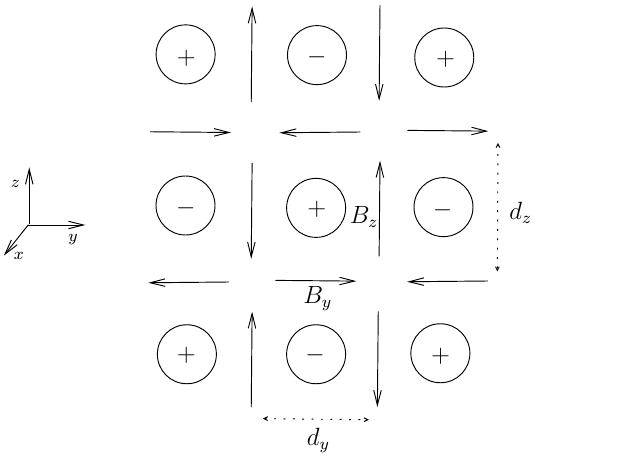}
	\caption{\small A finite-energy background made with staggered currents can create a pattern of oscillating magnetic field in two directions.
	}
	\label{currents}
\end{figure}

 We take a more realistic background with an inhomogeneous, alternating magnetic field.  This can be achieved in different ways. For example, we can take
\beq
a_x = C \cos{(k_z z)} \cos{(k_y y)}
\label{trig}
\eeq
with $d_z= \frac{\pi}{k_z}$, $d_y= \frac{\pi}{k_y}$
where
\beq
B_z = - C k_y \cos{(k_z z)} \sin{(k_y y)} \ , \qquad \quad  B_y = C k_z \sin{(k_z z)} \cos{(k_y y)} \ .
\eeq
The background is thus schematically of the type in Figure \ref{currents}. 
Note that this is not a solution of the equations of motion by itself. There must be alternating external currents forced to stay in the $x$ direction.  In any case it is a finite-energy configuration in field space, and it satisfies by definition the electric Bianchi identity. 
In proximity of  the planes $ y = \frac{\pi}{k} (\frac{1}{2} + n)$ we have the maximum 
$
B_z \simeq -C k_y \cos{(k_z z)} \left(1 + k_y^2(y-\pi (\frac{1}{2} + n))^2\right)
$
We take the monopole radius $R_M \ll \frac{1}{k_z}$ and we can consider the pair-production problem on the oscillatory backgrounds
\begin{align}
&	B_z \simeq (-1)^n C k_y \cos{(k_z z)} \ , \qquad \quad  &  y = \frac{\pi}{k_y} \Big(\frac{1}{2} + n\Big) \label{Bz}  \ , \nn \\ 
&	B_y \simeq (-1)^n C k_y \cos{(k_y y)} \ , \qquad \quad &   z = \frac{\pi}{k_z} \Big(\frac{1}{2} + n\Big) \ .
\end{align}
We want $R_M \ll d_y,d_z$ but also $ d_y,d_z \ll \lambda_{conf}$, where $ \lambda_{conf}$ is the length scale of confinement we will compute later. We want to send $B_z$ to critical value, while keeping $B_y$ below critical. For this we consider $d_z \geq d_y$.

We focus on the region where the magnetic field is maximal, say $x=y=\mathrm{const}$, and consider the $z,\tau$ plane, where
\beq
\tilde{a}_{\tau} = h(z) \qquad \quad B_z = \tilde{f}_{z \tau} = h'(z)
\label{BztA}
\eeq
We take 
\beq
h(z) = H \sin{(k_z z)} \qquad H = (-1)^{n}  \frac{C k_y}{k_z} \ .
\label{hsin}
\eeq
There is no background of this type that satisfies the Bianchi identity for $a_{\mu}$ everywhere. It can be regarded, as before, as a good approximation in a region of space to the staggered external currents discussed above (\ref{Bz}). For the moment we keep $h(z)$ generic; we take a generic periodic odd function with period $d_z$  which may arise from a distribution of currents like in Figure \ref{currents}.

\begin{figure}[!h]
	\centering
	\includegraphics[width=0.4\linewidth]{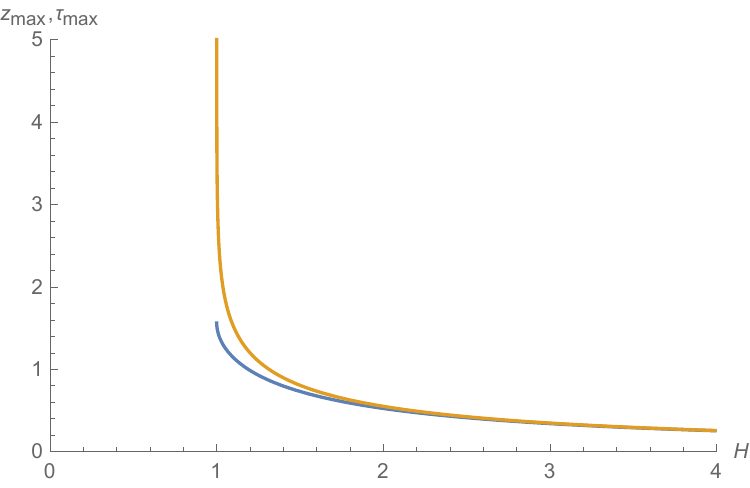} \quad 
	\includegraphics[width=0.55\linewidth]{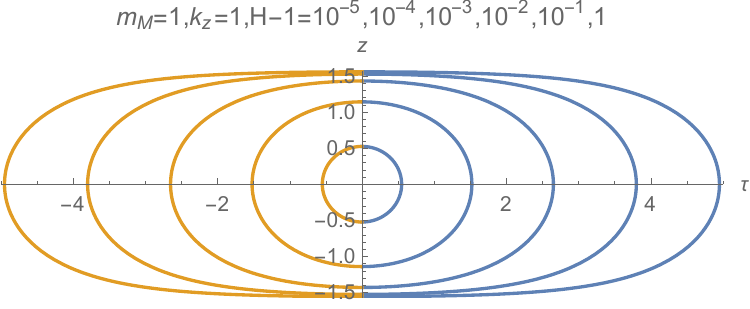}
	\caption{\small The instanton–monopole–loop solution (\ref{instmonloop}) for the oscillatory background for different $C$. Approaching the critical value the loop becomes infinitely elongated in the Euclidean time direction while remaining fixed at the scale of the spatial fluctuation. This effect is generic for any type of oscillation, not only the simple trigonometric one. We used $m_{M}=1$ and $k_z=1$.
	}
	\label{oscillatory}
\end{figure}

\begin{figure}[htp]
	\centering
	\includegraphics[width=0.42\linewidth]{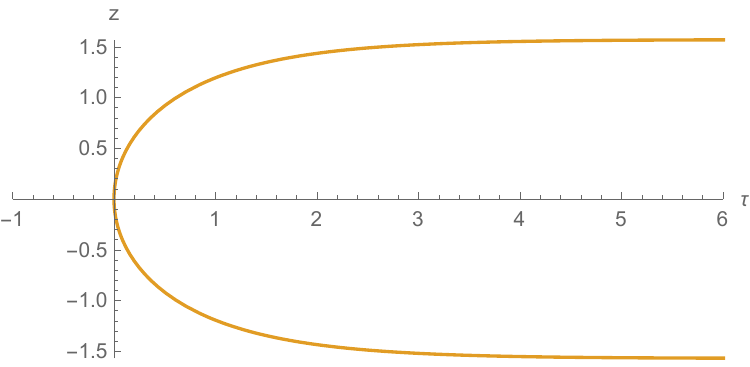} \qquad 
	\includegraphics[width=0.42\linewidth]{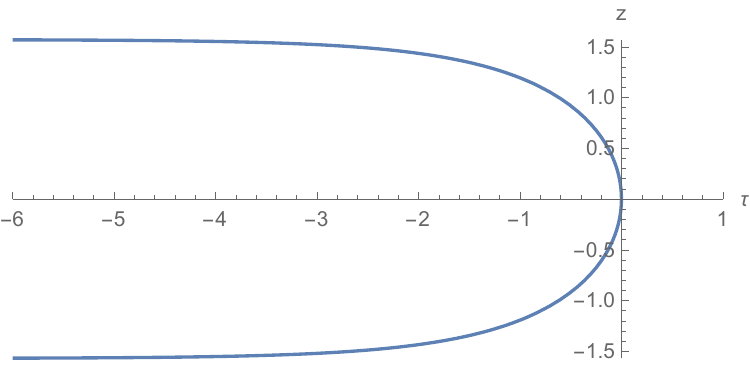}
	\caption{\small The deconfined bits of the monopole loop (\ref{taucrittrig}) at the critical value $B_{cr}$. We used $m_{M}=1$ and $k_z=1$. The position in $\tau$ can be arbitrary (the constant in (\ref{taucrittrig}) which here in the plot is set to $0$). 
	}
	\label{critical}
\end{figure}

We use the action  (\ref{worldlineaction}) with the background (\ref{BztA}) and, for convenience, we now parameterize the worldline with $l=z$ in
\beq
S_{E} = \int dz \left(m_{M} \sqrt{1+\tau'^2} + h(z) \tau' \right) \ .
\label{worldlineactionpz}
\eeq
The constant of motion from the invariance under $\tau$ translations is
\beq
m_{M} \frac{\tau'}{\sqrt{1+\tau'^2}} + h(z) = {\rm const }= h(0) = 0 \ .
\eeq
In this way we fixed the constant so that $\tau'(0) = 0$ which is the expected solution when $B_z(z)$ reaches the maximum on $z=0$, as in (\ref{Bz}). The solution in integral form is
\beq
\tau_{\pm}(z) = \tau_{\pm}(0) \mp \int_{0}^z ds \frac{h(s)}{\sqrt{m_{M}^2 -h(s)^2}} \ .
\label{tauint}
\eeq
To have a full loop we must combine the two $\pm$ solution with a proper choice 
\beq
\tau_{+}(0)-  \tau_{- }(0) = 2 \int_{0}^{z_{max}} ds \frac{h(s)}{\sqrt{m_{M}^2 -h(s)^2}} \ , \qquad h(z_{max}) = m_{M} \ .  
\eeq
where $ z_{max}$ is the  maximal value of $z$ reached by the loop. 
The critical value to a have a deconfined monopole string is reached when the integral (\ref{tauint}) has a divergence, and that is 
\beq
z_{max} =d_z \ , \qquad h_{cr}(d_z) = m_M \ .
\label{fcrit}
\eeq
In the specific case above (\ref{hsin}) with simple trigonometric spatial
oscillations, there is an analytic solution \cite{Dunne:2005sx}, which in our parametrization choice is given by
\beq
\tau_{\pm}(z) = \pm \frac{1}{k_z} \log{\Big(\sqrt{2} H \cos{(k_z z)} + \sqrt{2 m_{M}^2-H^2(\cos{(2k_z z)} - 1)}\Big)} + {\rm const_{\pm} } \ .
\label{instmonloop}
\eeq
where to close the loop we need a relation between the two constants
\beq
{\rm const_{+}} - {\rm const_{-}} = -  \frac{2}{k_z} \log{\Big(\sqrt{2} H \cos{\Big(\frac{1}{2} \arccos{\Big(\frac{H^2-2m_M^2}{H^2}\Big)} } \Big) \Big)}
\eeq
The plot for different values of $B$ is given in Figure \ref{oscillatory}.  
At the critical value the monopole loop acquires an additional almost zero mode and the monopole “bits” effectively deconfine.   The approach to the critical limit exhibits the elongation in the $\tau$ direction of the monopole loop. In this case we   see clearly the mechanism of gradual  elongation of the monopole loop near $B_{cr}$. 
The critical value is for
\beq
H_{cr} = C_{cr} \frac{k_y}{k_z}= m_{M} 
\eeq
 which is consistent with the general formula (\ref{fcrit}). The solution for the deconfined monopole string is
\beq
\tau_{cr,\pm}(z) = \pm \frac{1}{k_z}\log (\cos{(k_zz)}) +{\rm const } \ ,
\qquad \quad  S_E= 2 \frac{m_{M}}{k_z} \ .
\label{taucrittrig}
\eeq
We plot them in Figure \ref{critical}. These are the analogues of the monopole and antimonopole in the generalized Polyakov mechanism of confinement. The semi-infinite strings have infinite mass but this is canceled by the magnetic field contribution for $B = B_{cr}$. Also the two semi-infinite strings have opposite orientation so do not contribute to the magnetic charge, at least at large distance. Effectively these solutions behave like segments of monopole string between the two plates, thus effectively like a $3$D monopole and antimonopole. 

There is another type of critical magnetic field which is related to the W-boson spectrum, and not directly to the monopole; we refer to this as $B_{cr, UV}$ to distinguish it from the previous critical field. In a constant magnetic field the  spectrum of a charged spin-$1$ particle like the W-boson is \cite{Nielsen:1978rm} 
\beq
E^2 =  (2 n +1) B + 2  k   B + m_W^2 \ ,\qquad k=- 1,0,+1 \ ,\qquad n \in \mathbb{N} \ ,
\eeq
where the Zeeman term splits  $2  k   B $ the three spin states  of the vector and    $n=0$ is the ground state.
For $B >  B_{cr, UV} = m_W^2  $ we  have a tachyonic instability for $n=0$, $k=-1$.  We want to stay below this critical value, so 
\beq
B_{cr} <  m_W^2 
\label{condcrUV}
\eeq
is another condition we need to impose.

\begin{figure}[h]
	\centering
	\includegraphics[width=0.45\linewidth]{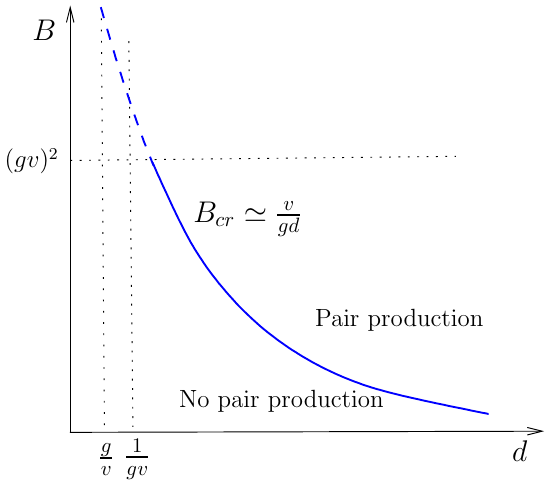} 
	\caption{\small  Critical line in the $B$-$d$ plane and the other conditions of Eq. (\ref{conditions}).
	}
	\label{Bd}
\end{figure}

Let us consider $d_y \sim d_z =d $ of the same  the order of magnitude, and $B_{cr} \simeq \frac{v }{g d_z}$. 
 All the requirements,  the weak-coupling condition $g \ll 1$, and also with the semiclassical monopole-loop approximation $ \frac{vd}{g} \gg 1 $,  and (\ref{condcrUV}), which is the strongest,  can be satisfied  if we take
 \beq
 \frac{1}{ vd} <g^3 \ll g \ll  \frac{1}{g}\ .
 \label{conditions}
 \eeq
 So for small $g$ and sufficiently large $v d$ it is possible to satisfy all the conditions. We summarize the various conditions in Figure \ref{Bd}.

 \begin{figure}[htp]
 	\centering
	\includegraphics[width=0.65\linewidth]{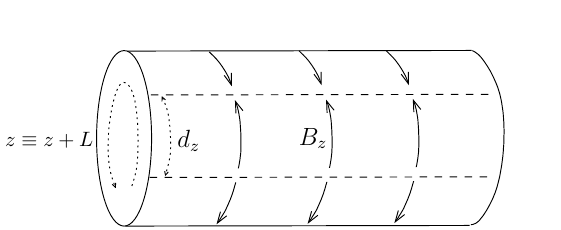} 
 	\caption{\small Compactification with a magnetic background. We show the direction $x$ (or $\tau$) at the value $y=d_y(\frac{1}{2} + n)$ where there is critical pair production. 
 	}
 	\label{compact}
 \end{figure}
 Now we consider the instanton-monopole computation for a background of the type of Figure \ref{currents} at the critical value.  We still consider compactification by $L$, so we want to repeat the same $3$D computation, requiring all length scales to remain larger than $L$. The instanton-monopole or antimonopole, now is one of the deconfined bits of the monopole loop in Figure \ref{compact}, and the action is half of the monopole loop computed at $B=B_{cr}$. It is an infinite, not closed monopole string.   Zero modes of a single monopole are parameterized by $\mathbb{R}^2 \times S^1$.
  The measure is now
 \beq
  d\mu^{(3)} = \xi^{(3)} dx dy d\tau \ ,\qquad \quad\xi^{(3)} \propto N \Big(\sqrt{S_{inst}} m_{UV}\Big)^3 \frac{1}{m_{W}} \frac{1}{d_y} e^{-S_{inst}} \ ,
 \qquad N = \frac{L}{d_z} \ , \eeq
 with
 \beq
 S_{inst} = S_{M} = \frac{\beta \alpha v d_z}{ g}   \ , \qquad m_{UV} \simeq m_W \ .
 \label{sm}
 \eeq
  $\alpha$ is the same coefficient defined earlier, and $\beta$ is a coefficient that depends on the background, $\beta = \frac{2}{\pi}$ from (\ref{taucrittrig}) for a simple trigonometric function (\ref{trig}).  
 One difference from before is that $S_{M}$ no longer increases with $L$ - this is the important aspect that blocks the exponential dependence on $L$ we had earlier. Another different is in the presence of the factor $N$ counts the multiplicity of layers, and thus of monopoles. From a $3$D perspective these are $N$ distinct, but equivalent, species of monopoles.
 Thus the average separation in the monopole–antimonopole gas is
 \beq
 \lambda_{gas} \simeq \frac{ d_y^{1/3}}{L^{1/3} v^{7/6} d_z^{1/6} g^{1/6}} e^{ \frac{\beta \alpha  v d_z}{3 g} }
 \label{lgas3d}
 \eeq
 and the squared mass of the dual photon is
 \beq
 M_{\varphi}^2 = \frac{1}{\lambda_{conf}^2} \simeq \frac{2 \xi^{(3)}d_z^2}{e^2 L^2} \ .
 \label{massqvphi}
 \eeq
 The factor $\left(\frac{d_z}{L}\right)^2$ is due to the dilution of the charge - the monopole charge in 3D is now given by the bit of the string that extends in the $z$ segment between the walls, and that is a fraction $\frac{d_z}{L}$  of the full compactified circle. 
 The confinement length scale is thus
 \beq
 \lambda_{conf} \simeq \frac{g^{3/4}  d_y^{1/2}}{ v^{7/4} d_z^{5/4}} e^{ \frac{\beta \alpha  v d_z}{2 g} } 
 \label{lconf3d}
 \eeq
 
 The $L$ dependence of $\lambda_{gas}$ in (\ref{lgas3d}) is now only in the prefactor and, most importantly, $\lambda_{gas}$ decreases with $L$. At a certain point we will have
 $
 \lambda_{gas} \simeq L
$
 which happens at
 \beq
 L^* \simeq  \frac{ d_y^{1/4}}{ v^{7/8} d_z^{1/8} g^{1/8}} e^{ \frac{\beta \alpha  v d_z}{4 g} }  \ .
 \eeq
 At this scale we must have a transition between $3$D and $4$D descriptions, as the average monopole-antimonopole separation becomes of the same order as the compactification scale. This means we can no longer use the previous $3$D computation, but it is also a good thing: the confinement mechanism becomes a $4$D effect and does not disappear in the decompactification limit.
 
 The $L$-dependence of $\lambda_{gas}$ and $\lambda_{conf}$. In this we keep $d$ fixed and change $L$. For $L < d$ there is no magnetic field background and the length scales grow exponentially as in (\ref{scaleswithoutwalls}). The Dirac monopole walls start to appear for $L \geq d$ and we have to follow (\ref{lgas3d}),(\ref{lconf3d}).   When $\lambda_{gas}$ becomes of order $L$ the effect becomes essentially $4$D.
 

 Now we consider the full $4$D analogue of the Polyakov computation. We consider  $L \gg d$ and it could also be $\infty$. The instanton measure is
 \beq
 d\mu^{(4)} = \xi^{(4)} dx dy dz d\tau \qquad \quad \xi^{(4)} \propto \Big(\sqrt{S_{inst}} m_{UV}\Big)^3 \frac{1}{m_{W}} \frac{1}{d_z d_y} e^{-S_{inst}}
 \eeq
 with (\ref{sm}). The relation with the $3$D measure, in case $L$ is finite, is $d\mu^{(4)} = d\mu^{(3)} L$.
 The $4$D theory electromagnetic duality is
 \beq
 \frac{1}{g} f_{\mu\nu} = \frac{g}{2} \epsilon_{\mu\nu\rho\sigma} \tilde{f}_{\rho\sigma} \ ,
 \eeq
 with $f= da$ and $\tilde{f} = d \tilde{a}$. The interaction with monopoles is most conveniently written with the dual gauge field   $\tilde{a}_{\mu}$. The deconfined monopole string worldline interacts with the dual gauge fields as
 \beq
 \int \tilde{a}_{\mu} d x^{\mu} \simeq \tilde{a}_{z} d_z \ ,
 \label{approxsegment}
 \eeq
 where we used the fact that, for large spatial distances with respect to $d$, the contributions from the two semi-infinite strings cancel, since they have opposite orientation. 
The effective action in the dual-photon formalism, and in the dilute gas approximation is
 \beq
 S_E \simeq \int d^4x   \frac{g^2}{4} \tilde{f}^2 + \sum_k d_z \int dxdyd\tau   \xi^{(4)} \left(e^{i \int \tilde{a}_{\mu} d x^{\mu}} + e^{ - i \int \tilde{a}_{\mu} d x^{\mu}}\right)  \ .
 \label{dualeffective}
 \eeq  
 In the small field expansion, and using the approximation (\ref{approxsegment}), at large distance we get the quadratic action
 \beq
 S_E \simeq \int d^4x   \left( \frac{g^2}{4} \tilde{f}^2 +   2 \xi^{(4)} d_z^2 \tilde{a}_{z}^2    \right) \ .
 \label{dualeffectiveq}
  \eeq  
   The result for the mass of the dual photon
  \beq
  M_{\tilde{a}_z}^2 =   \frac{2\xi^{(4)} d_z^2}{g^2} \ .
  \label{masstA}
  \eeq
 When we compactify $z$ on radius $L$, the dual gauge field in $2+1$, the scalar $\varphi$ is related to the holonomy of  dual gauge field in $3+1$, and in this case to the component $\tilde{a}_z$ 
 \beq
 \varphi = \int_0^L dz \tilde{a}_z \simeq L \tilde{a}_z
 \eeq
  The result from (\ref{dualeffectiveq}) is thus the same as in (\ref{massqvphi})
 \beq
 M_{\tilde{a}_z}^2 = M_{\varphi}^2  \ .
 \eeq
 This mechanism gives a mass only to the component $\tilde{a}_z$ of the dual photon, which is the scalar field dual to the photon in $2+1$. It is thus an anisotropic mass term.
 To give mass to other components of the dual photon we need deconfined monopole bits in more than one direction and for this we will need to consider a different background.  Gauge field mass from Wilson loop is also discussed in \cite{Aharony:2008an}, and the mechanism is used in holographic QCD for example to give mass to the quarks.

 The length scales for the instanton gas and confinement are
 \beq
 \lambda_{gas} \simeq \frac{1}{( \xi^{(4)})^{1/4}} \simeq   \frac{ d_y^{1/4}} {g^{1/8} v^{7/8} d_z^{1/8}} e^{ \frac{ \beta \alpha v d}{4g}}  \ ,  \qquad \quad   \lambda_{conf} \simeq \frac{1}{M_{\tilde{a}_z }}  \simeq \frac{g^{3/4} d_y^{1/2}} { v^{7/4} d_z^{5/4}} e^{\frac{ \beta \alpha v d}{2g}}  \ .
 \eeq
 Again, the approximation works if the various length scales are related as
 \beq
 \lambda_{conf}  
 \gg   \lambda_{gas}  \gg d \gg \frac{1}{v g} \ ,
 \eeq
 and these can  be satisfied  for weak coupling $g < 1$ and sufficiently large $d v $.

If we take  $d_y = d_z = d $ we can reach criticality  for both type of monopole-instantons in the $z$ and $y$ direction at the same time.
   Each one will gives mass to a different component of the dual gauge field.
  \beq
 M_{\tilde{a}_z }^2 = M_{\tilde{a}_y }^2 = \frac{2 \xi^{(4)}d^2}{g^2} \simeq  \frac{ v^{7/2} d^{3/2}}{g^{3/2}}  e^{- \frac{ \beta \alpha v d}{g}} \ .
 \eeq
  Two components are enough in $4$ dimensions to give a mass gap to the entire dual photon, although still anisotropic.

  This configuration discussed before is a finite-energy background, although it is not a stationary solution of the theory. A related model in which this configuration of currents could be constructed as a solution of the equations of motion can be built with an extension of the model using superconducting strings \cite{Witten:1984eb} to source the magnetic field. So we consider the model (\ref{action4D}) with an extra $U(1)$ with coupling $g'$ gauge field and two more complex scalar fields. The fields and charges are as follows
\beq
\begin{tabular}{c|c|c}
	& $SU(2)$, \ $g$ & $U(1)$, \ $g'$ \\
	\hline
$\Phi$	& {\rm Adj} & 0 \\
	\hline
$\chi_1$	& (.) &  1 \\
	\hline
$\chi_2$	& {\bf 2} &  0
\end{tabular}
\eeq
and the potential is 
\beq
V = \lambda \Big(  \tr \Phi^2 -  \frac{v^2}{2} \Big)^2 + k_1 \big(|\chi_1|^2-w_1^2 \big)^2  - k_2 \big( w_2^2 - |\chi_1|^2\big) |\chi_2|^2  \ .
\eeq
This model can accommodate both magnetic monopoles and superconducting vortex strings.
We consider the case $w_1 \gg v$. So the $U(1)$ breaks at  high energy  by the $\chi_1 =w_1$ condensate and Abrikosov-Nielsen-Olesen string form at this scale; later we see the reason for  that. We want $ w_2 > w_1$ so that $\chi_1=0$ in the bulk but in the center of the string $\chi_2$ is tachyonic and may condense if $k_2$ is sufficiently large. This is the standard superconducting string model scenario. The $\chi_2$ condensate may carry superconducting currents for the unbroken gauge field, and thus magnetic field around the string. 
The other field $\Phi$ then also breaks $SU(2) \to U(1)$ and creates magnetic monopoles. For $g' < k_1$ the Abrikosov lattice  of  type II strings is stable. Let us denote $d$ as the distance between the strings. We use  the condensate to turn on superconducting currents, and thus the magnetic field of the unbroken $U(1) $ field between the strings. Criticality is reached when $\frac{I}{d} \sim B \sim \frac{m_{M}}{d} $ where $I$ is the electric current on the strings. So we need a current of order  $I \sim \frac{v}{g}$ to reach the critical value for the magnetic field.  If  $w_1 \ll v$, the $U(1)$ breaking scale is much higher than the $SU(2)$ and thus a current $\frac{v}{g}$ would have a very small impact on the vortex and the stability of the Abrikosov lattice. This shows that it is in principle possible to achieve a stable  background, solution to the equation of motion and without UV divergences in the energy,  with critical magnetic field for the monopole.

In this work we discussed the critical case $B = B_{cr}$. 
The case $B>B_{cr}$ may still lead to confinement, as the liberation of monopole antimonopole is real as the vacuum is unstable to monopole pair production. 
The analysis of confinement in the transient period where monopole are pair produced is a more difficult problem that involves real time dependence. 

\subsection*{Acknowledgments}

The work is supported by the INFN special research project
grant ``GAST'' (Gauge and String Theories).


\begin{thebibliography}{99}
	
	\bibitem{Polyakov:1976fu}
	A.~M.~Polyakov,
	``Quark Confinement and Topology of gauge Groups,''
	Nucl. Phys. B \textbf{120} (1977), 429-458
	
	\bibitem{Shifman:2008ja}
	M.~Shifman and M.~Unsal,
	``QCD-like Theories on R(3) x S(1): A Smooth Journey from Small to Large r(S(1)) with Double-Trace Deformations,''
	Phys. Rev. D \textbf{78} (2008), 065004
	[arXiv:0802.1232 [hep-th]].
	
	\bibitem{Polyakov:2004vp}
	A.~M.~Polyakov,
	``Confinement and liberation,''
	[arXiv: hep-th/0407209 [hep-th]].
	
	\bibitem{Bolognesi:2011rq}
	S.~Bolognesi and K.~Lee,
	``1/4 BPS String Junctions and $N^3$ Problem in 6-dim (2,0) Superconformal Theories,''
	Phys. Rev. D \textbf{84} (2011), 126018
	[arXiv:1105.5073 [hep-th]].
	
	\bibitem{Nguyen:2025voy}
	M.~Nguyen and M.~{\"U}nsal,
	``Self-dual monopole loops, instantons and confinement,''
	[arXiv:2509.09625 [hep-th]].
	
	\bibitem{Auzzi:2008zd}
	R.~Auzzi, S.~Bolognesi, M.~Shifman and A.~Yung,
	``Confinement and Localization on Domain Walls,''
	Phys. Rev. D \textbf{79} (2009), 045016
	[arXiv:0807.1908 [hep-th]].
	
	\bibitem{Dvali:2007nm}
	G.~Dvali, H.~B.~Nielsen and N.~Tetradis,
	``Localization of gauge fields and monopole tunnelling,''
	Phys. Rev. D \textbf{77} (2008), 085005
	[arXiv:0710.5051 [hep-th]].
	
	\bibitem{Brezin:1970xf}
	E.~Brezin and C.~Itzykson,
	``Pair production in vacuum by an alternating field,''
	Phys.\ Rev.\ D {\bf 2} (1970) 1191.
	
	\bibitem{Schutzhold:2008pz}
	R.~Schutzhold, H.~Gies and G.~Dunne,
	``Dynamically assisted Schwinger mechanism,''
	Phys.\ Rev.\ Lett.\  {\bf 101} (2008) 130404
	[arXiv:0807.0754 [hep-th]].
	
	\bibitem{Affleck:1981ag}
	I.~K.~Affleck and N.~S.~Manton,
	``Monopole Pair Production in a Magnetic Field,''
	Nucl. Phys. B \textbf{194} (1982), 38-64
	
	\bibitem{Affleck:1981bma}
	I.~K.~Affleck, O.~Alvarez and N.~S.~Manton,
	``Pair Production at Strong Coupling in Weak External Fields,''
	Nucl. Phys. B \textbf{197} (1982), 509-519
	
	\bibitem{Bolognesi:2012gr}
	S.~Bolognesi, F.~Kiefer and E.~Rabinovici,
	``Comments on Critical Electric and Magnetic Fields from Holography,''
	JHEP {\bf 1301} (2013) 174
	[arXiv:1210.4170 [hep-th]].
	
	\bibitem{tHooft:1974kcl}
	G.~'t Hooft,
	``Magnetic Monopoles in Unified Gauge Theories,''
	Nucl. Phys. B \textbf{79} (1974), 276-284
	
	\bibitem{Polyakov:1974ek}
	A.~M.~Polyakov,
	``Particle Spectrum in Quantum Field Theory,''
	JETP Lett. \textbf{20} (1974), 194-195
	
 
	
	\bibitem{Dunne:2005sx}
	G.~V.~Dunne and C.~Schubert,
	``Worldline instantons and pair production in inhomogeneous fields,''
	Phys. Rev. D \textbf{72} (2005), 105004
	[arXiv: hep-th/0507174 [hep-th]].
	
	\bibitem{Bolognesi:2016qme}
	S.~Bolognesi, E.~Rabinovici and G.~Tallarita,
	``String pair production in non homogeneous backgrounds,''
	JHEP \textbf{04} (2016), 174
	[arXiv:1601.04758 [hep-th]].
	
\bibitem{Nielsen:1978rm}
N.~K.~Nielsen and P.~Olesen,
``An Unstable Yang-Mills Field Mode,''
Nucl. Phys. B \textbf{144} (1978), 376-396
	
	\bibitem{Shifman:2012zz}
	M.~Shifman,
	``Advanced topics in quantum field theory.: A lecture course,''
	Cambridge Univ. Press, 2012,
	
	\bibitem{Aharony:2008an}
	O.~Aharony and D.~Kutasov,
	``Holographic Duals of Long Open Strings,''
	Phys. Rev. D \textbf{78} (2008), 026005
	[arXiv:0803.3547 [hep-th]].
	
	\bibitem{Witten:1984eb}
	E.~Witten,
	``Superconducting Strings,''
	Nucl. Phys. B \textbf{249} (1985), 557-592
	
\end{thebibliography}
\end{document}